\documentclass[letterpaper,conference]{IEEEtran}
\pdfoutput=1
\usepackage{calc,cite}
\usepackage{amsfonts,amsmath,amssymb,graphics,psfrag,epsfig,graphics}
\usepackage[mathcal]{euscript}  
\usepackage{hyperref}
\hypersetup{colorlinks = true, citecolor = blue}

\DeclareMathOperator{\sinc}{sinc}
\newcommand{\defeq}{\triangleq}

\newcommand{\SNR}{\mathrm{SNR}}
\newcommand{\diff}{\mathrm{d}}
\newcommand{\conv}{\ast}
\newcommand{\T}{\mathrm{T}}
\newcommand{\kd}{\tilde{k}}
\newcommand{\bKd}{\tilde{\bK}}
\newcommand{\zd}{\tilde{z}}
\newcommand{\bzd}{\tilde{\bz}}
\newcommand{\yd}{\tilde{y}}
\newcommand{\byd}{\tilde{\by}}
\newcommand{\xd}{\tilde{x}}
\newcommand{\bvspace}{\bv}
\newcommand{\cgt}{\dagger}
\newcommand{\Tnyq}{T_0}

\newcommand{\bH}{{\mathbf{H}}}

\newcommand{\bK}{{\mathbf{K}}}

\newcommand{\bv}{{\mathbf{v}}}

\newcommand{\bx}{{\mathbf{x}}}

\newcommand{\by}{{\mathbf{y}}}

\newcommand{\bz}{{\mathbf{z}}}

\DeclareMathAlphabet{\mathbsf}{OT1}{cmss}{bx}{n}
\DeclareMathAlphabet{\mathssf}{OT1}{cmss}{m}{sl}
\DeclareMathAlphabet{\mathcsf}{OT1}{cmss}{sbc}{n}

\DeclareSymbolFont{bsfletters}{OT1}{cmss}{bx}{n}  
\DeclareSymbolFont{ssfletters}{OT1}{cmss}{m}{n}
\DeclareMathSymbol{\bsfGamma}{0}{bsfletters}{'000}
\DeclareMathSymbol{\ssfGamma}{0}{ssfletters}{'000}
\DeclareMathSymbol{\bsfDelta}{0}{bsfletters}{'001}
\DeclareMathSymbol{\ssfDelta}{0}{ssfletters}{'001}
\DeclareMathSymbol{\bsfTheta}{0}{bsfletters}{'002}
\DeclareMathSymbol{\ssfTheta}{0}{ssfletters}{'002}
\DeclareMathSymbol{\bsfLambda}{0}{bsfletters}{'003}
\DeclareMathSymbol{\ssfLambda}{0}{ssfletters}{'003}
\DeclareMathSymbol{\bsfXi}{0}{bsfletters}{'004}
\DeclareMathSymbol{\ssfXi}{0}{ssfletters}{'004}
\DeclareMathSymbol{\bsfPi}{0}{bsfletters}{'005}
\DeclareMathSymbol{\ssfPi}{0}{ssfletters}{'005}
\DeclareMathSymbol{\bsfSigma}{0}{bsfletters}{'006}
\DeclareMathSymbol{\ssfSigma}{0}{ssfletters}{'006}
\DeclareMathSymbol{\bsfUpsilon}{0}{bsfletters}{'007}
\DeclareMathSymbol{\ssfUpsilon}{0}{ssfletters}{'007}
\DeclareMathSymbol{\bsfPhi}{0}{bsfletters}{'010}
\DeclareMathSymbol{\ssfPhi}{0}{ssfletters}{'010}
\DeclareMathSymbol{\bsfPsi}{0}{bsfletters}{'011}
\DeclareMathSymbol{\ssfPsi}{0}{ssfletters}{'011}
\DeclareMathSymbol{\bsfOmega}{0}{bsfletters}{'012}
\DeclareMathSymbol{\ssfOmega}{0}{ssfletters}{'012}

\begin{document}

\title{Super-Nyquist  Rateless Coding for \\
  Intersymbol Interference Channels}
\author{\authorblockN{Uri Erez}
\authorblockA{Tel Aviv University, Ramat Aviv, Israel\\
Email: uri@eng.tau.ac.il}
\and
\authorblockN{Gregory W.~Wornell}
\authorblockA{Dept. EECS, MIT, Cambridge, MA\\
Email: gww@mit.edu}
}

\maketitle \renewcommand{\thefootnote}{} 
\footnotetext[1]{This work was presented, in part, at the 2012
  International Zurich Seminar in Communications (IZS).  This work was
  supported in part by ONR under MURI Grant No.~N00014-07-1-0738,
  by AFOSR under Grant No.~FA9550-11-1-0183, and by the Israel Science
  Foundation under Grant No.~1557/12.}
\renewcommand{\thefootnote}{\arabic{footnote}}

\begin{abstract}
A rateless transmission architecture is developed for communication
over Gaussian intersymbol interference channels, based on the concept
of super-Nyquist (SNQ) signaling.  In such systems, the signaling rate
is chosen significantly higher than the Nyquist rate of the system.
We show that such signaling, when used in conjunction with good
``off-the-shelf'' base codes, simple linear redundancy, and minimum
mean-square error decision feedback equalization, results in
capacity-approaching, low-complexity rateless codes for the
time-varying intersymbol-interference channel.  Constructions for both
single-input / single-output (SISO) and multi-input / multi-output
(MIMO) ISI channels are developed.
\end{abstract}

\section{Introduction}

In traditional digital communication, achieving high throughput when
the channel state allows is accomplished by selecting high-order
signal constellations.  However, an alternative approach, originally
proposed several decades ago \cite{Mazo75}, exploits super-Nyquist
(SNQ) (equivalently, faster-than-Nyquist) signaling.  In SNQ
signaling, the symbols are taken from a fixed constellation, typically
BPSK or QPSK, independent of the transmission rate.  Higher rates are
achieved by increasing the signaling rate---i.e., the rate at which
the symbols are modulated onto the bandlimited pulse shape---beyond
the Nyquist rate.  Thus, in SNQ systems, the signaling rate is
decoupled from the transmission bandwidth, and can greatly exceed the
transmission bandwidth.

Because SNQ modulation introduces ISI, it necessitates the use of
equalization, which traditionally made it unappealing for early
applications; see, e.g., \cite{Foschini84}.  In this paper, however,
we establish that SNQ signaling has some particularly valuable
properties for communication over Gaussian intersymbol interference
(ISI) channels where the transmitter knows neither the channel impulse
response nor the maximal rate that my be supported by the channel.  In
particular, we establish the somewhat surprising result that the use
of SNQ signaling allows for highly efficient joint design of the
physical and link layers.  Indeed, from such signaling we develop a
rich family of low-complexity, capacity-approaching rateless codes for
scalar ISI channels, which have natural extensions to vector ones.

\section{System and Channel Model}

We consider a linear dispersive Gaussian channel for which the complex
baseband channel output takes the form
\begin{equation*}
y(t) = h(t) \conv x(t)+z(t),
\end{equation*}
where $z(t)$ is additive white Gaussian noise (AWGN) with one-sided
power spectral density $N_0$, and where $x(t)$ is the input, which is
subject to a power constraint $\mathbb{E}\{|x(t)|^2 \} \leq P$ and
bandwidth constraint $W$.  The associated white-input capacity of the
channel is
\begin{equation}
C_\mathrm{[b/s]} 
= \int_{-W/2}^{W/2}\log\left(1+ \frac{P |H(f)|^2 }{N_0 W}\right)\,\diff f.
\label{eq:cap_ct}
\end{equation}

We consider pulse-amplitude modulation, whereby
\begin{equation}
x(t)=\sum_n s[n] \, g(t-nT),
\label{eq:pulse_modulation}
\end{equation}
where $T$ is the symbol duration and $\Tnyq=1/W$ is the Nyquist
sampling time.  The associated ``over-signaling'' ratio is thus
$L=\Tnyq/T$.  After matched filtering and sampling at the
symbol rate, the equivalent discrete-time channel is
\begin{equation}
y[n] = k[n] \conv s[n] + z[n],
\label{dt_channel}
\end{equation}
where $k[n] =k(nT)$, where
$k(t) = h^*(-t) \conv h(t) \conv g(-t)^* \conv g(t)$,
and where
\begin{align*}
S_{zz}(e^{j 2 \pi f }) &=\frac{N_0}{2}K(e^{j 2 \pi f}) \\
                       &=\frac{N_0}{2T} \sum_i 
|H(f/T+i/T)|^2|G(f/T+i/T)|^2.
\end{align*}
The pulse shape is required to be limited to system bandwidth $W$,
i.e., $g(t)$ satisfies $G(f)=0$ for $|f|>W/2$.  To simplify our
development, we largely restrict our attention to the case
\begin{equation}
g(t)=\sinc(t/\Tnyq),\text{ with }\sinc(u) \defeq \sin(\pi u)/(\pi u).
\label{eq:sinc}
\end{equation}

Taking the symbols $s[n]$ to be independent identically-distributed
(i.i.d.) circularly symmetric complex Gaussian with power $P/L$
results in a (proper) Gaussian random input signal $x(t)$ with power
$P$.  It follows that the capacity of the discrete-time channel
\eqref{dt_channel} is
\begin{align}
&C_\mathrm{[b/SNQ\ symbol]} \notag\\
&\ = \int_{-1/2}^{1/2}\log\left(1+ \frac{(P/L) \cdot  K(e^{j 2 \pi f
    })}{N_0}  \right)\,\diff f  \notag \\
&\ = {T}  \int_{-1/2T}^{1/2T}\!\log\left(1\!+\! \frac{P {\sum_i
      |H(f\!+\!i/T)|^2|G(f\!+\!i/T)|^2}}{\Tnyq N_0}   \right)\,\diff f \notag  \\
             & = {T}  \int_{-1/2T}^{1/2T}\log\left(1+ \frac{P
    {|H(f)|^2|G(f)|^2}}{\Tnyq N_0}   \right)\,\diff f,
\label{eq:cap_snq_nonideal}
\end{align}
where the last equality follows from the fact that $g(t)$ is bandlimited.

Note that for sinc modulation \eqref{eq:sinc}, $x(t)$ has a flat power
spectrum over the bandwidth $W$, and the modulation achieves the
white-input capacity \eqref{eq:cap_ct} for any $L$, i.e.,
\eqref{eq:cap_snq_nonideal} specializes to
\begin{align}
C_\mathrm{[b/SNQ\ symbol]} 
& = \frac{1}{LW}
\int_{-W/2}^{W/2}\log\left(1+ \frac{P |H(f)|^2 }{N_0 W}\right)\,\diff f
\notag \\
& = \frac{T}{\Tnyq} \frac{1}{W} C_\mathrm{[b/s]}. 
\label{eq:cap_snq}
\end{align}

\section{Linear SNQ Rateless Coding}
\label{sec:lin-snq}

Consider now packetized transmission where the packet size is large
but otherwise plays no role in the analysis.  We consider a simplified
model where the channel response experienced throughout transmission
of the $m$\/th packet, $m=1,\ldots,M$, is linear time-invariant (LTI)
but the impulse response, which we denote by $h_m(t)$, may vary from
packet to packet.  The channel input-output relation for the
transmission of the $m$\/th packet is therefore
\begin{equation}
y_m[n] = s_m[n] \conv k_m[n] + z_m[n],
\label{eq:y_m}
\end{equation}
where $k_m[n]=k_m(nT)$ and $k_m(t)= h_m^*(-t) \conv h_m(t) \conv
g^*(-t) \conv g(t)$.  Assuming discrete-time white-input transmission
for all packets, it follows from \eqref{eq:cap_snq}, that the mutual
information (in b/SNQ~symbol) corresponding to each packet is
\begin{align*}
C_{m \mathrm{ [b/SNQ \ symbol]}} 
&= \int_{-1/2}^{1/2}\log\left(1+ \frac{P K_m(e^{2 \pi f}) }{N_0
    L}\right)\, \diff f \notag \\
&= \frac{1}{L} \frac{1}{W} \int_{-W/2}^{W/2}\log\left(1+ \frac{P
    |H_m(f)|^2 }{N_0 W}\right)\,\diff f,
\end{align*}
where the second equality holds for ideal sinc modulation.
Upon receiving a set $\mathcal{S}\subset \{1,\ldots,M\}$ of packets, the aggregate mutual information is thus
\begin{align}
C(\mathcal{S})=\sum_{m \in \mathcal{S}}C_m.
\label{eq:vr}
\end{align}

Our aim is to design a low complexity coding and modulation scheme
that (simultaneously) approaches $C(\mathcal{S})$ for all sets
$\mathcal{S}\subset \{1,\ldots,M\}$ without requiring the transmitter
to have knowledge of the capacities $C_m$. Rather, for any given
chosen target rate $R$, and no knowledge of the channel, transmission
should be successful whenever $C(\mathcal{S})>R$ holds for the
received set of packets $\mathcal{S}$.

We proceed to describe the proposed linear rateless SNQ
construction. All the signals $s_m[n]$ are obtained from a single
coded stream $s[n]$ according to
\begin{equation}
s_m[n]=v_m[n]\,s[n],
\label{eq:dithered_inputs}
\end{equation}
where $v_m[n]$ are sequences to be specified. 
The transmitted signal corresponding to packet $m$ is thus
\begin{align*}
x_m(t)=\sum_i s[i]\, v_m[i] \,g(t-iT).
\end{align*}

Provided we choose the sequences $v_m[n]$ so that the transmitted
signals $x_m(t)$ are statistically independent and the $s_m[n]$ are
white circularly-symmetric complex Gaussian processes, the mutual
information corresponding to each packet remains $C_m$ and furthermore
the aggregate mutual information from the receipt of multiple packets
is the sum of the individual ones.

A simple means to achieve this is by taking $v_m[n]=e^{-j 2 \pi 
  mn/L}$.   Accordingly, we define
\begin{equation} 
x_m[n] \triangleq x_m(nT) =\sum_i s[i]\,e^{-j 2 \pi im/L}   \, g[n-i],
\label{eq:dft-x}
\end{equation}
where $g[n] = g(nT) = \sinc(n/L)$, with $g(t)$ as in \eqref{eq:sinc}.
Using \eqref{eq:dft-x}, we see that the transmit signals
``shifted-back in frequency'' in this case are
\begin{align*}
\xd_m[n] \triangleq e^{j 2 \pi mn/L}\,x_m[n] =\sum_i s[i]\, g_m[n-i],
\end{align*}
where $g_m[n] =e^{j 2 \pi mn/L}\, g[n]$.

Clearly, requiring that the signals $\{x_m(t)\}$ be mutually
independent is equivalent to requiring that the associated
discrete-time signals $\{x_m[n]\}$ be.  Furthermore, the latter holds
if and only if $\{\xd_m[n]\}$ are mutually independent.  Therefore, it
suffices to verify the last condition.  Since the signals $\xd_m[n]$
are jointly Gaussian and stationary, they are independent if their
cross-spectra
\begin{align*}
&S_{\xd_{m_1}\xd_{m_2}}(e^{j 2 \pi f}) \notag\\
&\quad = S_{ss}(e^{j 2 \pi f})\, G_{m_1}(e^{j 2 \pi f})
 \, G^*_{m_1}(e^{j 2 \pi f}) \\
&\quad =\frac{1}{T^2}S_{ss}(e^{j 2 \pi f})\, G\left(\frac{f+m_1/L}{T} \bmod\frac{1}{T}\right) \notag \\
&\qquad\qquad\qquad\qquad\quad {} \cdot G^*\left(\frac{f+m_2/L}{T} \bmod \frac{1}{T} \right)
\end{align*}
vanish.  Since $G(f/T)$ occupies no more than $1/L$ of the SNQ
frequency band, it follows that there is no overlap between the
frequency responses $G\left((f+m/L)/T \bmod 1/T \right)$ for different
values of $m$, and hence $S_{x_{m_1}x_{m_2}}(e^{j 2 \pi f})$ indeed
vanishes for $m_1 \neq m_2$. 

\subsection{Receiver Architecture}
\label{sec:mmse_dfe}

A low-complexity receiver architecture suffices to approach the
associated information-theoretic limits.  In particular, specializing
\eqref{eq:y_m} to the case \eqref{eq:dithered_inputs} with our choice
$v_m[n]=e^{-j 2 \pi mn/L}$, the equivalent ``shifted back" channel
model is
\begin{equation} 
\yd_m[n] \triangleq y_m[n]\, e^{j 2 \pi mn/L} = \kd_m[n] \conv s[n] + \zd_m[n],
\label{eq:y_m3*}
\end{equation}
where $\kd_m[n]=k_m[n]\,e^{j 2 \pi mn/L}$.  Since these channels do
not overlap in frequency, they may be added without loss, resulting in
the effective scalar ISI channel
\begin{align*}
\yd[n] &= \sum_{m \in \mathcal{S}} y_m[n]\, e^{j 2 \pi mn/L} \\
       &= s[n] \conv \left( \sum_{m \in \mathcal{S}} \kd_m[n] \right) + \sum_{m \in \mathcal{S}}\zd_m[n].
\end{align*}
For such ISI channels, the unbiased MMSE decision-feedback equalizer
(DFE) is an information-lossless receiver structure
\cite{GuessVaranasiIT,Forneyallerton04}.  In particular, to approach
capacity, one can use Guess-Vanarasi interleaving
\cite{GuessVaranasiIT} and a single (fixed-rate) base code designed
for an AWGN channel. In essence, every symbol is replaced by a
different codeword and thus the DFE decision device acts on codewords
rather than symbols.  

\section{MIMO-SNQ: Extending SNQ to MIMO systems}

A straightforward extension of the preceding architecture to a
multi-input multi-output (MIMO) system, which we term MIMO-SNQ, is as
follows.  The particular channel model of interest is
\begin{equation*}
\by_m(t) = \bH_m(t) \conv \bx_m(t) + \bz_m(t),
\end{equation*}
where there are $N_\mathrm{t}$ transmit and $N_\mathrm{r}$ receive
elements.  In turn, with input of the form
$\bx_m(t) = \sum_n \bx_m[n]\, g(t-nT)$,
the associated discrete-time channel,
after applying a matrix matched filter, is
\begin{align*}
\by_m[n] &= \bK_m[n] \conv \bx_m[n] + \bz_m[n],
\end{align*}
where $\bK_m[n] = \bK_m(nT)$ with $\bK_m(t) = \bH^\cgt_m(-t) \conv
\bH_m(t) \conv g^*(-t) \conv g(t)$.

We employ a single stream transmission architecture based on the
application of time-varying DFT beamforming to a scalar signal
$s[n]$. Specifically, we assume that the transmitted signal
corresponding to packet $m$ is formed as
$\bx_m[n]= s[n]\, \bv_m[n]$,
where
$\bv_m[n]=\bvspace[n]\, e^{-j 2 \pi mn/L}$
with
\begin{align*}
\bvspace[n]=\begin{bmatrix} 1 & e^{-j 2 \pi n/N_\mathrm{t}} &
\cdots &
e^{-j 2 \pi (N_\mathrm{t}-1)n/N_\mathrm{t})} \end{bmatrix}^\T.
\end{align*}

The effective received signal for packet $m$ is, after
frequency shifting [cf.\ \eqref{eq:y_m3*}],
\begin{equation}
\byd_m[n] = \by_m[n]\,e^{j2\pi mn/L} = \sum_l \bKd_m[l]\, \bvspace[n-l]\, s[n-l] + \bzd_m[n],
\label{eq:y_m6}
\end{equation}
where $\bKd_m[n] = \bK_m[n]\,e^{j2\pi mn/L}$.  Note that the effective
channel \eqref{eq:y_m6} is a periodically varying MIMO-ISI channel
with period $N_\mathrm{t}$. We refer to each of the $N_\mathrm{t}$
induced substreams as ``phases.''  As in the SISO case, we may employ
a DFE at the receiver, but due to the time-varying nature of the
effective channel, for each phase a different set of $N_\mathrm{r}$
feedforward filters is applied to the channel output vector
sequence. Thus, the equalizer is also periodic with period
$N_\mathrm{t}$.

Note the covariance matrix of the transmitted vector for our SNQ
modulation is white.  We conclude that for this modulation, the
transmitted signal is white over all degrees of freedom as long as
the oversignaling rate satisfies $L \geq N_\mathrm{t} M$.  However,
this doesn't guarantee capacity can be achieved.  In particular, we
associate with each of the $N_\mathrm{t}$ ``phases'' a
signal-to-interference-plus-noise ratio (SINR) value corresponding to
the associated DFE slicer input.  Equivalently, we may associate with
each such phase a corresponding capacity.  Hence, while the sum of the
per-phase capacities equals the white-input capacity of the MIMO
channel, the per-phase capacities are in general not equal.  Moreover,
since the variation is unknown to the transmitter, in a Guess-Varanasi
transmission architecture, a fixed code rate is used, and thus the
achievable rate is determined by the minimum of the per-phase
capacities.

It is worth emphasizing that this SINR variation across phases is
analogous to the SINR variation across streams in a V-BLAST system, in
which independently coded streams are sent over the antennas
\cite{vblast}.  For this reason, V-BLAST serves as a useful benchmark
with which to compare the performance of SNQ modulation.

\subsection{Parallel channels}
\label{sec:par}

In some cases, MIMO-SNQ is strictly capacity achieving.  For example,
consider the special case of $N$ parallel ISI channels.  This
model is essentially equivalent to using SNQ modulation for
transmission over a block-varying ISI channel as considered in
Section~\ref{sec:lin-snq}, with the $N$ parallel channels replacing
the $L$ consecutive blocks of the SISO ISI channel. As we have shown
that SNQ modulation is an optimal scheme in such a scenario, it
follows that MIMO-SNQ is optimal for the case of parallel channels.

\subsection{Channels Without Temporal ISI}

As another class of channels of interest, consider the special case in
which there is no temporal ISI and only inter-channel interference
(ICI) is present, i.e., $\bH_m(t)=\bH_m$.  

While for diagonal $\bH_m$ MIMO-SNQ is capacity achieving, there exist
other $\bH_m$ for which MIMO-SNQ achieves zero rate.  For example, if
$\bH_m$ is the (rank-one) matrix of all 1's, there will exist an SNQ
Nyquist substream that experiences a zero-capacity channel since a
vector of all 1's is orthogonal to $\bv[n]$ for $n\neq l
N_\mathrm{t}$, all $l$. Hence, MIMO-SNQ achieves zero rate, while
V-BLAST achieves a strictly positive rate.  However, for all but such
pathological $\bH_m$, MIMO-SNQ supports a rate that grows with SNR.
By contrast, it is well know that for all rank-one channel matrices
V-BLAST performance is interference-limited, i.e., is bounded with
increasing SNR.

More generally, when the channel matrix $\bH_m$ is drawn from a random
ensemble, its performance is on average never worse that V-BLAST, and
has significant advantages, particularly when keeping in mind that in
V-BLAST an ordering is forced in the detection process, while MIMO-SNQ
requires no such ordering since the scheme is inherently more
symmetric.  

\begin{figure}
\begin{center}
\leavevmode
\includegraphics[width=3.5in]{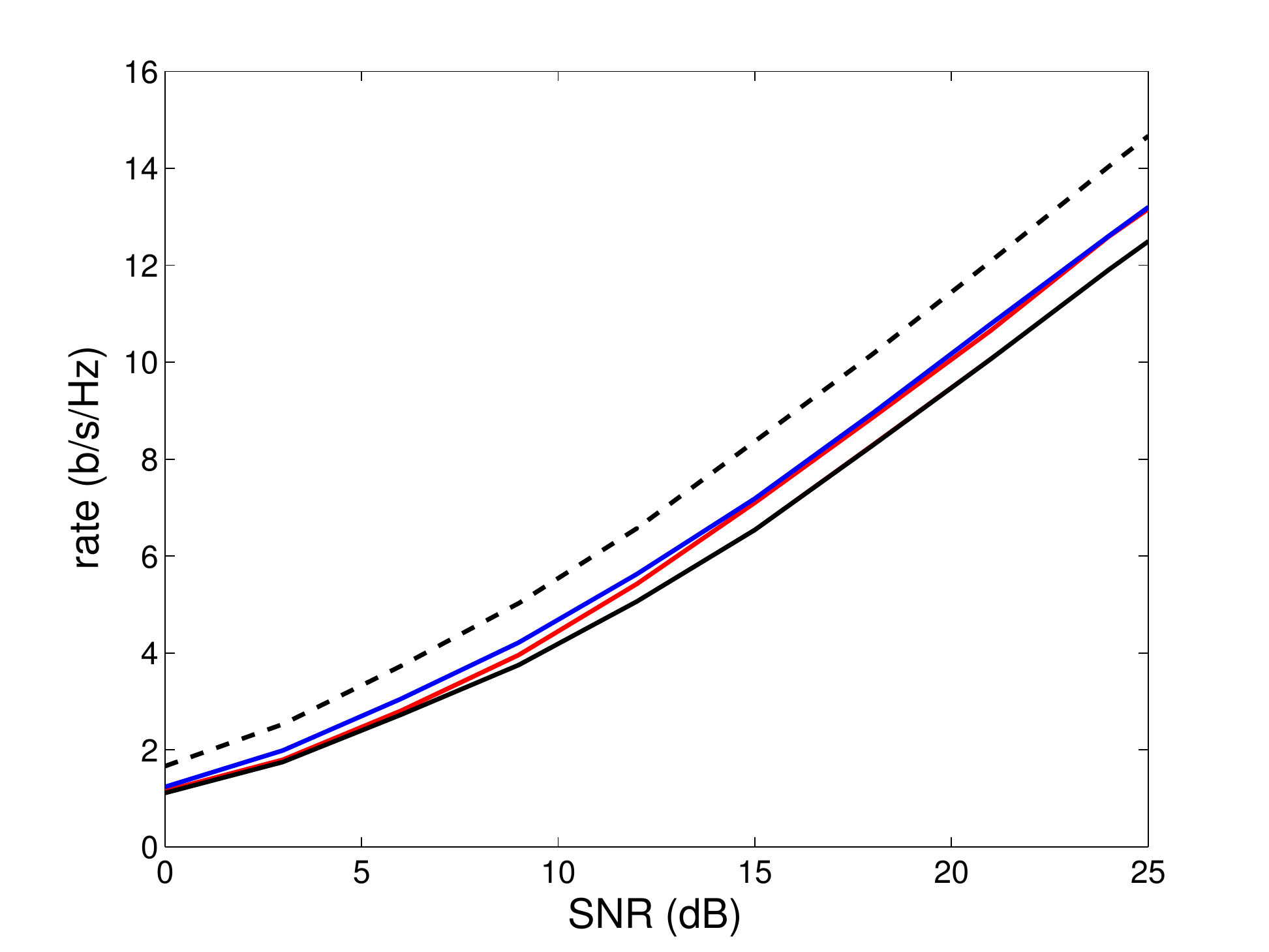}
\caption{Average MIMO-SNQ performance over random $2\times2$ MIMO
  channels without temporal ISI.  The successively higher solid curves
  correspond to V-BLAST with fixed-order decoding, MIMO-SNQ, and
  V-BLAST with optimum-order decoding, and the dashed curve indicates
  capacity.\label{Fig:vblast}}
\end{center}
\end{figure}

We consider the average throughput for an ensemble of $\bH_m$ with
i.i.d.\ circularly symmetric complex Gaussian entries. The resulting
average throughput of MIMO-SNQ is depicted in Fig.~\ref{Fig:vblast},
along with that for both fixed- and optimized-decoding-order V-BLAST
As the plot reflects, the performance of MIMO-SNQ modulation lies in
between the two and approaches the latter at high SNR.  

We can also relate MIMO-SNQ performance to that of D-BLAST
\cite{Foschini96}.  In particular, as is well known, D-BLAST can
achieve capacity, but to do so requires a base code designed for
time-varying channels.  In that sense, MIMO-SNQ can also achieve
capacity provided practical such base codes exist.  However, when we
require that a communication architecture has the property that the
base code sees an AWGN channel, Fig.~\ref{Fig:vblast} reflects that
MIMO-SNQ can perform as well as V-BLAST with an optimized decoding
order.  In constrast, when the same constraint is imposed on D-BLAST,
the result is V-BLAST with a fixed decoding order.

\subsection{Spatio-Temporal ISI Channels}

\begin{figure}
\begin{center}
\leavevmode
\includegraphics[width=3.5in]{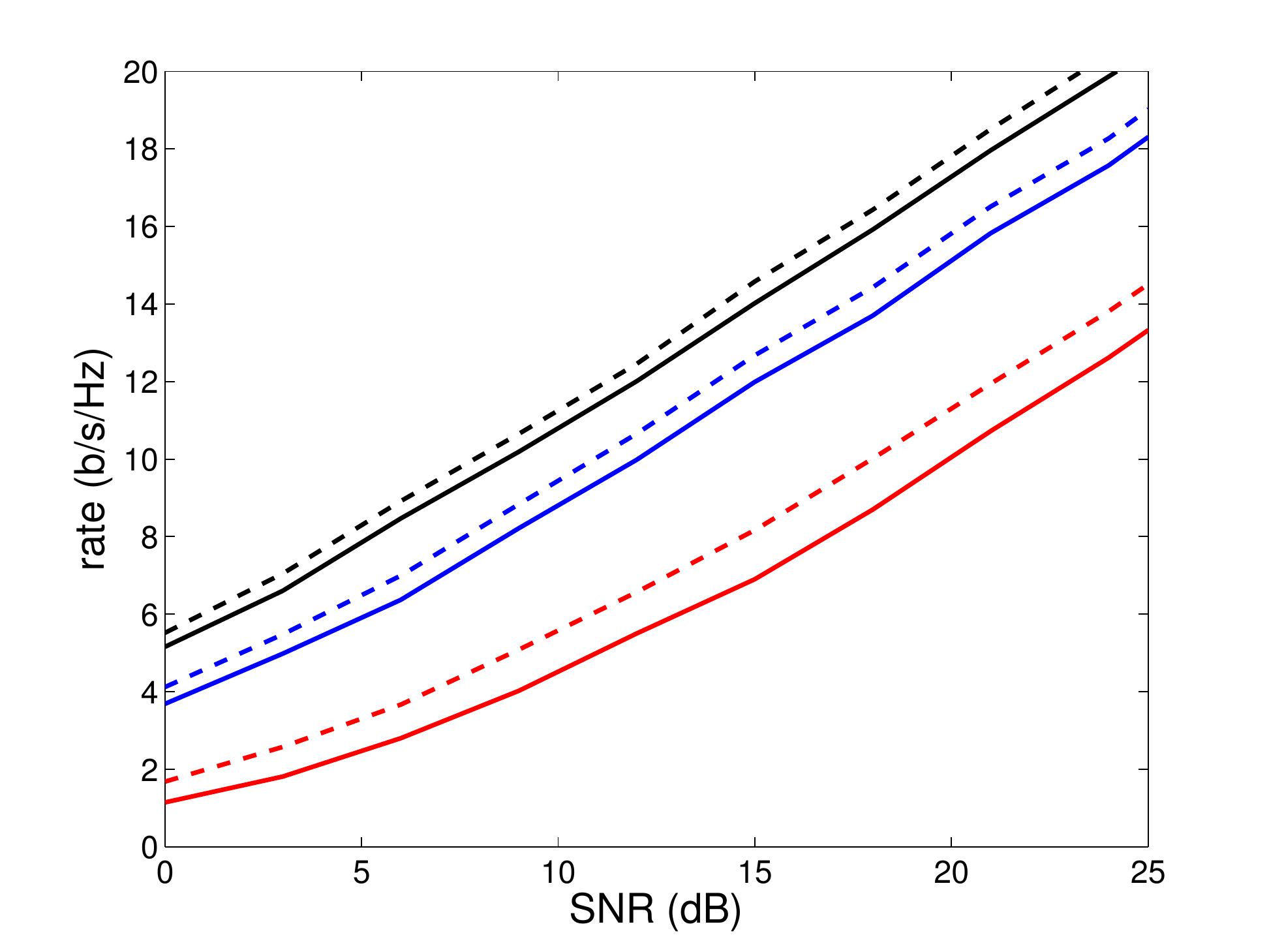}
\caption{Average MIMO-SNQ performance over random $2\times2$ MIMO
  channel with spatio-temporal uncorrelated Gaussian scattering. The
  successively higher solid curves correspond to 1, 5, and 10 taps of
  ISI, respectively, and the associated dashed curves indicate the
  corresponding capacities.\label{Fig:randomISI}}
\end{center}
\end{figure}

When there is also temporal ISI, MIMO-SNQ is even more attractive, as
we next illustrate.  We now consider a random MIMO channel model where
the Nyquist-rate equivalent discrete-time matrix channel impulse
response $\bK_m[n]$ is of finite length and each Nyquist-rate tap is
drawn i.i.d.\ over spatial and time dimensions according to a
circularly-symmetric complex Gaussian
distribution. Fig.~\ref{Fig:randomISI} depicts the expected (averaged
over the ensemble) capacity of MIMO-SNQ for different channel lengths.
We observe that the gap-to-capacity decreases as the channel length
grows confirming that SNQ modulation is able to exploit the temporal
diversity afforded by the channel.

Moreover, such behavior is not specific to such i.i.d.\ ensembles.
Indeed, Table~\ref{table:uwa} describes the performance of MIMO-SNQ
for a typical $2 \times 2$ underwater acoustic communication channel
realization from the recent KAM-11 experiment.\footnote{The authors
  thank Qing He for providing this sample channel.}  Performance was
numerically evaluated for an oversignaling rate of $L=2$. In this
case, the equivalent discrete-time (Nyquist-rate) baseband channel
impulse responses are 100 taps long.  As the table reflects, MIMO-SNQ
is effectively capacity achieving for this channel, with a gap to
capacity of less than 0.5 dB.

\begin{table}
  \caption{MIMO-SNQ spectral efficiency (b/s/Hz) on a sample $2\times
    2$ underwater acoustic channel of length 100 taps \label{table:uwa}}
\begin{center}
\begin{tabular}{c|cccccccc}
 & \multicolumn{8}{c}{SNR (dB)} \\
  & 0 & 2 & 4 & 6 & 8 & 10 & 12 & 14\\
\hline
\text{Capacity} & 0.95 & 1.31 & 1.76 & 2.30 & 2.94 & 3.68 & 4.51
  & 5.41 \\
\text{SNQ} & 0.86 & 1.20 & 1.63 & 2.16 & 2.80 & 3.53 &
  4.37 & 5.29
\end{tabular}
\end{center}
\end{table}

\bibliographystyle{ieeetr}
\bibliography{mybib_allerton2011}

\end{document}